\documentclass[12pt,preprint]{aastex}
\usepackage{graphicx}
\usepackage{amssymb,amsmath}

\newcommand\beq{\begin{equation}}
\newcommand\eeq{\end{equation}}
\newcommand\lsim{\mathrel{\rlap{\lower4pt\hbox{\hskip1pt$\sim$}}
        \raise1pt\hbox{$<$}}}
\newcommand\gsim{\mathrel{\rlap{\lower4pt\hbox{\hskip1pt$\sim$}}
        \raise1pt\hbox{$>$}}}

\begin{document}
\title{Thermo-Resistive Instability of Hot Planetary Atmospheres}

\author{Kristen
  Menou\altaffilmark{1}}
 
\altaffiltext{1}{Department of Astronomy, Columbia University, 550
  West 120th Street, New York, NY 10027}

\begin{abstract}
The atmospheres of hot Jupiters and other strongly-forced exoplanets
are susceptible to a thermal instability in the presence of ohmic
dissipation, weak magnetic drag and strong winds. The instability
occurs in radiatively-dominated atmospheric regions when the ohmic
dissipation rate increases with temperature faster than the radiative
(cooling) rate. The instability domain covers a specific range of
atmospheric pressures and temperatures, typically $P \sim 3$--$300$
mbar and $T \sim 1500-2500$~K for hot Jupiters, which makes it a
candidate mechanism to explain the dayside thermal ``inversions''
inferred for a number of such exoplanets.  The instability is
suppressed by high levels of non-thermal photoionization, in possible
agreement with a recently established observational trend. We
highlight several shortcomings of the instability treatment presented
here. Understanding the emergence and outcome of the instability,
which should result in locally hotter atmospheres with stronger levels
of drag, will require global non-linear atmospheric models with
adequate MHD prescriptions.
\end{abstract}

\section{Introduction}

A variety of gaseous exoplanets with strongly-forced atmospheres have
been discovered by astronomers (e.g., Charbonneau 2009). These
exoplanets, exemplified by the hot Jupiter class, receive extreme
levels of irradiation from their stellar host and likely experience
permanent day- and night-side forcing conditions, from being
tidally-locked on very compact orbits. Observationally, two of the
most interesting trends emerging from studies of well-characterized,
transiting hot Jupiters have been the tendency for many such planets
to exhibit radius inflation, well above expectations from standard
planetary cooling models, and the inference that thermal excesses
(inversions) are present on the dayside of some of the
strongly-irradiated planets (see Deming \& Seager 2009; Baraffe et
al. 2010; Burrows \& Orton 2010; Winn 2010 for reviews).

It has recently been proposed that hot, strongly-forced exoplanet
atmospheres are the site of significant magnetic induction when fast,
weakly-ionized atmospheric winds cross the pre-existing planetary
magnetic field. This induction takes the form of magnetic drag acting
to brake atmospheric winds and associated ohmic heating in the
planetary atmosphere and interior (Batygin \& Stevenson 2010; Perna,
Menou \& Rauscher 2010a,b; Menou 2012). While other mechanisms have
been proposed, ohmic heating is currently one of the leading
contenders to explain the inflated radii of hot Jupiters (Batygin et
al. 2011; Laughlin et al. 2011; Menou 2012; Wu \& Lithwick 2012).

In this letter, we explore the possibility that thermal inversions in
hot Jupiter atmospheres are not caused by extra absorption of stellar
light at altitude, as they have traditionally been interpreted (Hubeny
et al. 2003; Fortney et al. 2006, 2008; Burrows et al. 2008;
Madhusudhan \& Seager 2010), but instead result from the same
induction mechanism that may explain radius inflation. In this
alternative interpretation, thermal inversions have their origin in a
thermo-resistive instability that affects radiatively-dominated
atmospheric regions under specific conditions of weak magnetic drag,
strong ohmic dissipation and fast winds.

\section{Thermo-Resistive Instability Criterion} \label{sec:linear}

Let us first establish the existence of a thermo-resistive instability
under physical conditions relevant to hot exoplanet atmospheres.

{\bf Energy Equation} -- The energy equation satisfied by an
atmosphere experiencing ohmic dissipation can be written

\begin{equation}
\rho C_p \frac{d T}{d t} = \rho C_p \left( \frac{\partial T}{\partial t} + {\bf v \cdot \nabla}T \right)= -{\bf \nabla \cdot F_{\rm rad}} + Q_{\rm ohm} = Q_{\rm rad} + Q_{\rm ohm}, \nonumber
\end{equation}
where $\rho$ is the mass density, $C_p$ is the specific heat at
constant pressure and $Q_{\rm ohm}$ is the volumic ohmic dissipation
rate. In the last expression, the divergence of the net radiative
flux, ${\bf F_{\rm rad}}$, has been rewritten as a net volumic radiative
heating rate.

In radiatively-dominated regions of an atmosphere, which is typically
the case at and above infrared photospheric levels in hot Jupiter
atmospheres (Seager et al. 2005, Cowan \& Agol 2011, Menou 2012; Perna
et al. 2012), the advective term (${\bf v \cdot \nabla}T$) can be
neglected by comparison to the net radiative term ($Q_{\rm
  rad}$). Furthermore, near radiative equilibrium, this net radiative
term can expanded into a first-order Taylor series around the local
radiative equilibrium temperature, $T_{\rm eq}$ (Goody \& Yung 1989),
so that the energy equation in a radiatively-dominated atmosphere
reduces to

\beq
\frac{\partial T}{\partial t} = \frac{T_{\rm eq} -T}{\tau_{\rm rad}} +\frac{Q_{\rm ohm}}{\rho C_p }, \label{eq:thermal}
\eeq  

where $\tau_{\rm rad}$ is the time to relax to radiative equilibrium
in this so-called Newtonian approximation. Ohmic dissipation provides
a strictly positive contribution, so that in steady-state ($\partial T
/\partial t =0$), one expects thermal balance through net radiative
cooling, at an atmospheric temperature $T > T_{\rm eq}$. This defines
the basic radiative-ohmic state on which we wish to perform a
stability analysis.

{\bf Magnetic Induction Formalism} -- We use the steady-state
axisymmetric magnetic induction framework introduced by Liu et
al. (2008) and further discussed in the hot Jupiter context by Batygin
\& Stevenson (2010), Perna et al. (2010a,b) and Menou (2012). In this
framework, a weakly-ionized zonal wind with velocity scale $V_\phi$
flows across the planetary magnetic field of surface strength
$B$. This induces an additional magnetic field in the atmosphere and,
in steady-state, this magnetic induction is balanced by resistive
diffusion and associated ohmic dissipation in the atmosphere.

Lacking detailed models for how electric currents flow in the
generally inhomogeneous atmospheres of hot Jupiters and other hot
exoplanets, we evaluate the heating term associated with ohmic
dissipation in Eq.~(\ref{eq:thermal}) simply as

\beq
H_{\rm ohm} \equiv \frac{Q_{\rm ohm}}{\rho C_p } =\frac{V^2_{\phi} B^2}{4 \pi \eta \rho C_p } = \frac{V^2_{\phi} }{ \tau_{\rm drag} C_p}, \label{eq:ohm}
\eeq 

where $\eta$ is the electric resistivity and $\tau_{\rm drag}=4 \pi
\eta \rho / B^2$ is the corresponding magnetic drag time. The local
electric resistivity of the atmospheric gas is evaluated as $\eta= 230
\sqrt{T} /x_e~{\rm cm^2~s^{-1}}$, where $x_e$ is the free electron
ionization fraction (e.g., Menou 2012). This expression for $Q_{\rm
  ohm}$, which emerges from simple dimensional analysis of the
induction equation (Menou 2012), is consistent with more detailed
calculations which find that the bulk of the ohmic dissipation is
located in the active weather layer where induction occurs (e.g.,
Batygin \& Stevenson 2010; Perna et al. 2010b; Wu \& Lithwick
2012). The exponential dependence of $\eta$ and $Q_{\rm ohm}$ with
temperature when thermal ionization dominates is at the origin of the
thermo-resistive instability.

{\bf Linear Instability Criterion} -- Let us denote by $C_{\rm rad}$
the first, radiative relaxation term on the right-hand side of
Eq.~(\ref{eq:thermal}). Radiative-ohmic equilibrium is satisfied when,
in Eq.~(\ref{eq:thermal}), the heating term $H_{\rm ohm}$ (defined in
Eq.~[\ref{eq:ohm}]) exactly balances the cooling term $C_{\rm rad}$
(with $T>T_{\rm eq}$). This radiative-ohmic thermal equilibrium may be
unstable, however, if

\beq
\frac{d H_{\rm ohm}}{dT} > - \frac{d C_{\rm rad}}{dT}= \frac{1}{\tau_{\rm rad}},
\eeq \label{eq:inequal}

so that a slight temperature increment would lead to a greater
increase in ohmic heating than in net radiative cooling. In the limit
of weak drag, $V_{\phi}$ changes so weakly with temperature, despite
the varying magnetic drag (e.g., Menou 2012 or Eq.~[\ref{eq:drag}]
below), that

\beq
\frac{d H_{\rm ohm}}{dT} \simeq \frac{d H_{\rm ohm}}{d \eta}\frac{d \eta}{dT} = -\frac{H_{\rm ohm}}{\eta}\frac{d \eta}{dT}
\eeq

To obtain a result in analytic form, we use here a simplified
expression for the ionization fraction, which only accounts for
potassium ionization (Balbus \& Hawley 2000),
\begin{eqnarray}
x_e\equiv\frac{n_e}{n_n}&=&6.47\times 10^{-13}\left(\frac{a_K}{10^{-7}}\right)^{1/2}
\left(\frac{T}{10^3}\right)^{3/4}\nonumber \\ 
&\times&\left(\frac{2.4\times 10^{15}}{n_n} 
\right)^{1/2}\frac{\exp(-25188/T)}{1.15\times 10^{-11}}\;,
\label{eq:xe}
\end{eqnarray}  
where $n_e$ and $n_n$ are the number densities of electrons and of
neutrals, respectively (in cm$^{-3}$), $a_K= 10^{-7}$ is the potassium
abundance (assumed to be solar), and $T$ is the temperature in K.
With this expression for $x_e$, we obtain
\begin{eqnarray}
\frac{d H_{\rm ohm}}{dT} &=& H_{\rm ohm} (2.5 \times 10^{-4} T_3^{-1} + 2.52 \times 10^{-2} T_3^{-2}) \nonumber \\
 & \simeq & 4.2 \times 10^{-9}~{\rm s}^{-1} \frac{V_{\phi,5}^2 B_1^2 T_3^{1/4}}{\rho^{3/2}} \exp(\frac{-25.188}{T_3}) \left[ 2.5 \times (10^{-4} T_3^{-1} +  10^{-2} T_3^{-2}) \right], \label{eq:ohmderiv}
\end{eqnarray} 

where we adopted $C_p = 1.4 \times 10^8$~erg~g$^{-1}$~K$^{-1}$, $T_3$
is the temperature $T$ in units of $1000$~K, $V_{\phi,5}$ is the zonal
velocity scale $V_\phi$ in units of km~s$^{-1}$ and $B_1$ is the
magnetic field strength in G.

The scaling in Eq.~(\ref{eq:ohmderiv}) reveals that instability is
favored at lower atmospheric densities (hence lower pressures), for
larger values of the magnetic field strength and the zonal wind
velocity scale, and at higher temperatures ($T \gsim 10^3$~K), with a
strong exponential $T$-dependence. For conditions representative of
the mbar pressure level on the dayside of a hot Jupiter such as
HD209458b, $T \sim 1500$~K, $\rho \sim 10^{-8}$~g~cm$^{-3}$, one
obtains $d H_{\rm ohm}/dT \sim 2.9 \times 10^{-3}$~s$^{-1}$ for
$B=10$~G and $V_\phi = 3.3$~km~s$^{-1}$. By comparison, the radiative
relaxation time $\tau_{\rm rad} \simeq 3 \times 10^3$~s at the mbar
level in such an atmosphere (e.g., Showman et al. 2008), which
indicates thermally unstable conditions since $d H_{\rm ohm}/dT >
1/\tau_{\rm rad}$. The instability would disappear, however, for lower
values of the magnetic field strength ($B \lsim 1$~G), the wind
velocity scale ($V_\phi \lsim 1$~km~s$^{-1}$) or the atmospheric
temperature ($T \lsim 1000$~K).

This derivation is useful to illustrate the potential for thermal
instability and to understand parameter dependencies but it is only
valid when magnetic drag is inefficient and when thermal ionization is
dominated by potassium. In order to establish more reliable domains of
instability, we now relax these assumptions and turn to a numerical
parameter space exploration.

\section{Thermo-Resistive Instability Domain}

{\bf Method} -- Relaxing the weak drag limit implies the need for a
magnetic drag law that specifies the magnitude of drag as a function
of atmospheric resistivity, $\eta$ and magnetic field strength,
$B$. In the interest of simplicity, we adopt here a parameterized drag
law inspired from the physical arguments developed in Menou (2012). We
evaluate the dragged zonal velocity scale, $V_\phi^d$, from the
drag-free value $V_\phi$ according to

\beq
V_\phi^d = \frac{V_\phi}{1+n_s (V_{\rm drag}/V_\phi)^{n_e}},
\eeq \label{eq:drag}

where $V_{\rm drag}=R_p/\tau_{\rm drag}$ is a characteristic drag
velocity evaluated from the drag time $\tau_{\rm drag}$ (previously
defined in Eq.~[\ref{eq:ohm}]) and the planetary radius, $R_p$. We use
the dragged velocity, $V_\phi^d$, in all our ohmic dissipation
calculations. This formulation satisfies the weak drag expectation
$V_{\phi}^d \to V_\phi$ at low temperatures (high resistivities) and
the strong drag exponential dropoff, $V_{\phi}^d \propto \eta$ (for
$n_e=1$), expected at high temperatures (e.g Menou 2012). The
additional dimensionless parameters $n_s$ and $n_e$, with default
values of unity, are used to control the threshold and the steepness
of the transition from the weak to the strong drag regimes. With this
drag law, the half-speed transition for a $P=60$~mbar pressure level
(near the thermal photosphere) occurs at a temperature $ T \sim
2250$~K for $B =3$~G, $ T \sim 1900$~K for $B =10$~G and $ T \sim
1650$~K for $B =30$~G, with a half-width $\Delta T \sim
200$--$300$~K. Adopting a weaker drag law, with $ n_s=0.3$ and
$n_e=1$, pushes these transitions to higher temperatures ($ T \sim
2500$, $2000$ and $1750$~K for $B =3$, $10$ and $30$~G).  While this
parameterized magnetic drag law is broadly consistent with the results
discussed in Menou (2012), it also clearly is one of our most
uncertain model ingredients.

We use here a more detailed, numerical solution to the Saha equation
for the thermal ionization fraction $x_e$ (described in Menou 2012),
rather than the analytic, potassium-only formulation discussed in the
previous section. This allows us to explore more consistently the low
resistivity and high temperature ($T > 1700$~K) regime which is
relevant to the hottest exoplanets. Finally, we adopt a zonal velocity
profile, with default scale $V_\phi=7$~km/s (before drag is applied),
that decreases exponentially with depth according to $\exp (-P_1)$,
where $P_1$ is the pressure in units of 1 bar. This velocity scale and
its rapid decline at high pressures are broadly consistent with the
results of Rauscher \& Menou (2012) and other published circulation
models on the properties of superrotating equatorial winds on hot
Jupiters.

We identify instability domains by numerically evaluating the ohmic
derivative term $d H_{\rm ohm}/dT$ and comparing it to the inverse of
the radiative relaxation time, $\tau_{\rm rad}$. Regions satisfying
the inequality in Eq.~(\ref{eq:inequal}) are considered unstable. We
estimate the radiative times on the basis of the work of Iro et
al. (2005) and Showman et al. (2008), which yield consistent results
for HD209458b. We note that the tabulated, planet-specific values
provided by Showman et al. (2008) approximately follow the simple
scaling $\tau_{\rm rad} \propto T^{-3}$ over a wide range of
pressures, as may be expected from the formal definition of $\tau_{\rm
  rad}$ (e.g., Goody \& Yung 1989). In the interest of generality, we
normalize our radiative times to the values found by Iro et
al. (2005), using fits provided in Heng et al. (2011), and at a each
pressure level, we scale the radiative time in proportion to $(T_{\rm
  iro}/T)^3$, where $T_{\rm iro}(P)$ is the corresponding temperature
value in the profile of Iro et al. (2005). While this procedure is not
expected to yield $\tau_{\rm rad} $ values more accurate than a factor
of a few, it is sufficient given other sources of uncertainty in our
modeling methodology.

{\bf Results} -- Figure~\ref{fig:one} presents dayside
temperature-pressure profiles for a prototypical hot Jupiter with
radius $R_p=10^{10}$~cm and surface gravity $g=900$~cm~s$^{-2}$. Fifty
logarithmically-spaced profiles are shown for values of the planetary
effective temperature from $T_{\rm eff}=1000$~K to $2500$~K. The
profiles are calculated with the one-dimensional radiative solution
presented by Guillot (2010), for a zero incidence angle and a dilution
factor $f = 0.5$ representing a dayside average. An internal heat flux
corresponding to $T_{\rm int}=150$~K is adopted. As shown by Guillot
(2010), even when adopting constant thermal and visible absorption
coefficients as is done here ($\kappa_{\rm th}=
10^{-2}$~cm$^2$~g$^{-1}$ and $\kappa_{\rm v}= 4 \times
10^{-3}$~cm$^2$~g$^{-1}$), these temperature-pressure profiles are
reasonably good solutions for hot Jupiter atmospheres over a large
range of insolation fluxes.  For reference, HD209458b has $T_{\rm eff}
\simeq 1420$~K which corresponds to the twentieth profile from the
left.

Thermo-resistive instability domains are also shown in color in
Fig.~\ref{fig:one} for a default magnetic drag law ($n_s=n_e=1$ in
Eq.~[\ref{eq:drag}]).  The green and red instability domains
correspond to a zonal velocity scale $V_\phi=7$~km~s$^{-1}$ and
magnetic field strengths of $B=10$ and $40$~G, respectively. The blue
instability domain corresponds to $V_\phi=10$~km~s$^{-1}$ and
$B=3$~G. For $V_\phi=7$~km~s$^{-1}$ and $B=3$~G no instability is
found. As expected, higher velocities promote instability. Larger
magnetic field strengths also promote instability and tend to shift
the instability domain to lower atmospheric temperatures.  Details of
the adopted T-P profiles (e.g., opacity coefficients, dilution factor
$f$) have only minor effects on the instability domains in the sense
that instability is primarily set by physical conditions which
correspond to specific regions of the T-P plane.

The boundaries of the instability domains, which can be extended in
pressure but are typically limited to a narrow range of temperatures,
can be understood as follows. To the left of each instability domain,
at low temperatures, resistivity varies too weakly with temperature
for instability to occur. At the bottom of each instability domain,
besides the rapid exponential drop of velocity with pressure, we find
that our assumption of a radiatively-dominated atmosphere breaks
down. We thus arbitrarily truncated the instability domain at the
point where $\tau_{\rm adv}= R_p/V_{\phi} = \tau_{\rm rad}$, with the
implicit assumption that strong advective cooling would likely
neutralize the instability. More stringent criterions ($\tau_{\rm rad}
\ll \tau_{\rm adv}$) would further reduce the instability domains to
pressure $\lsim 0.3$-$0.1$~bar in all our figures.  Finally, to the
right and at the top of each instability domain, the emergence of the
strong magnetic drag regime rapidly leads to $V_{\phi}^d \ll V_{\phi}$
and the effective disappearance of the instability.

Figure~\ref{fig:two} shows that a weaker drag law (with $ n_s=0.3$,
$n_e=1$ in Eq.~[\ref{eq:drag}]), for the same zonal velocity scale
$V_\phi=7$~km/s and magnetic field strength $B=10$~G, can result in a
significantly more extended instability domain (compare to the green
domain in Fig.~\ref{fig:one}), by pushing the transition to strong
drag to higher temperatures. Using $n_e>1$ in Eq.~(\ref{eq:drag}) to
steepen the transition to strong drag has only minor effects on the
instability domains.

Interestingly, the upper atmosphere of hot Jupiters and other
strongly-forced exoplanets are subject to strong photo-ionization
rates.  An examination of the arguments in section \S\ref{sec:linear}
shows that the addition of a temperature-independent contribution to
the ionization fraction does not impact the value of $d H_{\rm
  ohm}/dT$ or the instability criterion. However, the lower
resistivity due to the additional ionization results in stronger
magnetic drag, which is a limiting factor for the
instability. Figure~\ref{fig:three} shows the same instability domain
as in Fig.~\ref{fig:two}, when a constant ionization fraction
$x_e^+=10^{-8}$ is added everywhere in the atmosphere, to emulate the
effects of photoionization. This level of additional ionization
stabilizes the lowest-density, upper atmospheric regions, where
magnetic drag is the strongest. We find that the entire atmospheric
domain can be stabilized with the addition of a constant ionization
fraction $x_e^+ \sim 10^{-7}$.

Fortney et al. (2003) modeled sodium photoionization in the atmosphere
of HD209458b and found ionization fractions in excess of thermal
values at pressure levels $\lsim 30$--$100$~mbar, corresponding to
$x_e^+ \gsim 10^{-8}$ (for a solar abundance of sodium). This,
together with the results shown in Fig.~\ref{fig:three}, suggests that
plausible levels of photo-ionization may indeed control the occurrence
of the thermo-resistive instability in hot exoplanet atmospheres. At
particularly high photo-ionization levels (and possibly super-solar
atmospheric abundances), photo-ionization could entirely suppress the
instability. Provided one associates the thermo-resistive instability
with the observationally-inferred phenomenon of dayside temperature
inversions, these results would appear consistent, at least
qualitatively, with the observational trend established by Knutson et
al. (2010): hot Jupiters subject to the strongest photo-ionization
fluxes are the least likely to show such inversions.

\section{Discussion and Conclusion}

Our argument for the existence of a thermo-resistive instability in
the atmospheres of hot, strongly-forced exoplanets relies on a number
of assumptions which are not all well justified. Most importantly, the
induction formalism used here and in other studies of magnetic effects
in hot Jupiter atmospheres and interiors assumes axisymmetry and
steadiness. We already emphasized in Perna et al. (2010a) the clearly
non-axisymmetric properties of strongly-forced exoplanet atmospheres,
a situation that would only be aggravated by the onset of a localized
thermal instability (e.g., on the dayside).

\begin{figure*}[l]
\centering \includegraphics[scale=0.8]{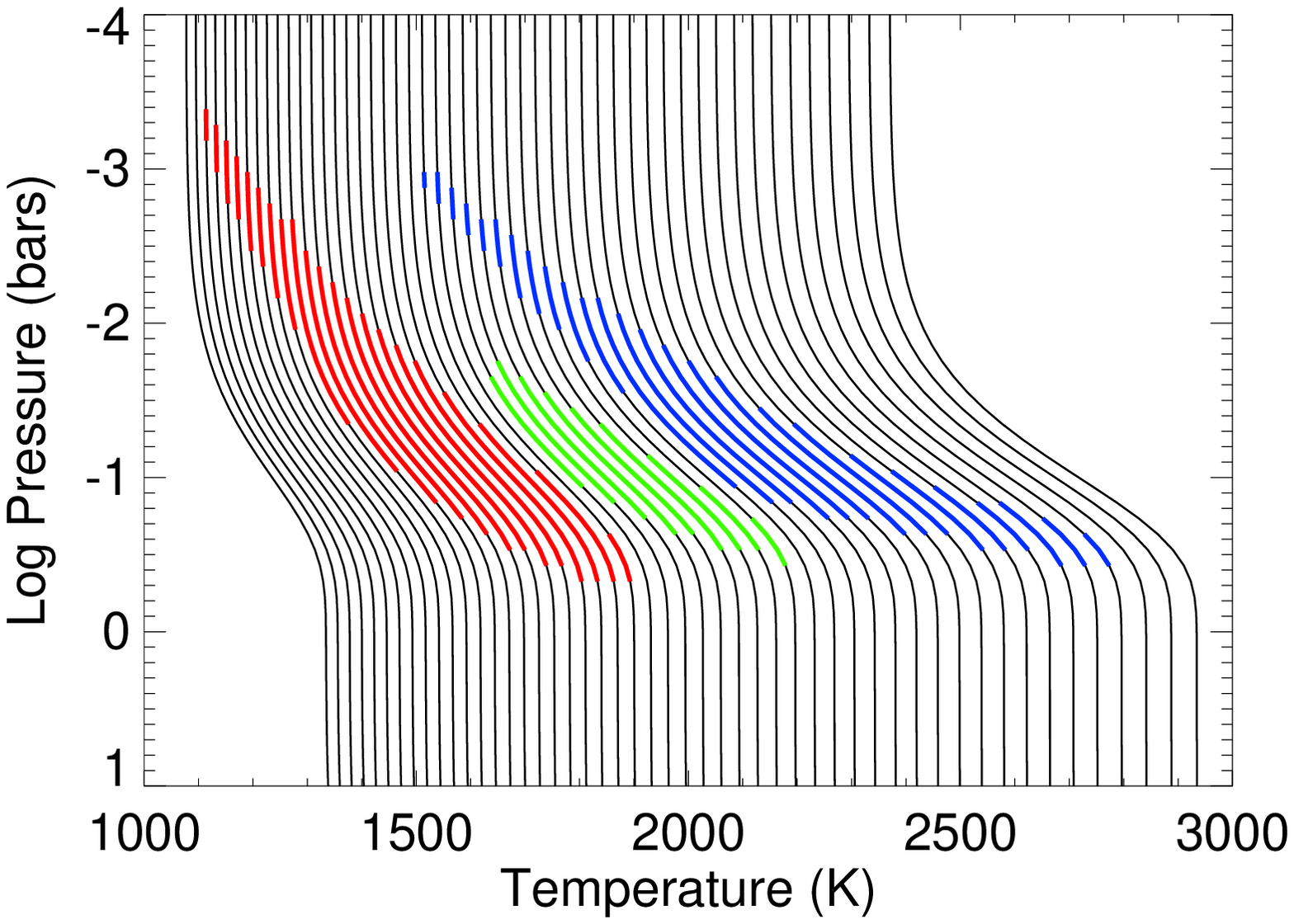}
\caption{Thermo-resistive instability domains superposed on
  temperature-pressure profiles representative of the dayside of a
  typical hot Jupiter, for the default magnetic drag law ($n_s=n_e=1$
  in Eq.~[\ref{eq:drag}]).  The green and red instability domains
  correspond to a zonal velocity scale $V_\phi=7$~km~s$^{-1}$ and
  magnetic field strengths $B=10$ and $40$~G, respectively. The blue
  instability domain corresponds to $V_\phi=10$~km~s$^{-1}$ and
  $B=3$~G.}
\label{fig:one}
\end{figure*}

Our study of an MHD-based instability with a steady-state induction
equation is another important shortcoming. Dimensional analysis of the
time-dependent axisymmetric induction equation (see Eq~[1] in Menou
2012) suggests that induced currents would lag changes in temperature
and resistivity by a resistive diffusion time, $\tau_{\rm diff} \sim
L^2/\eta$. Adopting a pressure scaleheight for the characteristic
lengthscale $L$, we estimate that $\tau_{\rm diff} < \tau_{\rm rad}$
for the green and red instability domains in Fig.~\ref{fig:one}, which
suggests that current-adjustment delays will not strongly affect the
instability development. At high temperatures and low densities,
however, around the blue instability domain and at higher temperatures
in Fig.~\ref{fig:one}, resistivities become small and $\tau_{\rm diff}
> \tau_{\rm rad}$, which may invalidate the implicit assumption of
instantaneous current adjustment made in our steady-state evaluation
of the ohmic dissipation term in Eq.~(\ref{eq:ohm}). A careful
consideration of this issue likely requires a full MHD treatment and
we shall simply note here that time-dependent current adjustments may
impact the thermo-resistive instability of the hottest exoplanet
atmospheres beyond the simple treatment adopted here.

Despite such limitations, it is tempting to associate the
thermo-resistive instability mechanism identified here with the
dayside thermal inversions inferred for a number of hot Jupiters. As
shown in Fig.~\ref{fig:one}, the instability domain is restricted to a
specific range of pressures and temperatures. For magnetic field
strengths $B \sim 3$-$30$~G, instability domains can match well the
upper atmospheric dayside conditions of moderately hot exoplanets such
as HD209458b, with low enough nightside temperatures $T \lsim 1100$~K
to suppress the instability. On cooler hot Jupiters such as HD189733b,
dayside conditions would barely achieve instability (see $T$-$P$
profiles in Showman et al. 2008 and Rauscher \& Menou 2012). By
contrast, very hot exoplanets with $T_{\rm eff} > 1500$-$2000$~K may
be too hot on their daysides for instability, and even if their
nightsides were to meet the temperature requirements, the instability
may be suppressed because of the very effective wind drag exerted on
the dayside or possibly strong deviations from radiative equilibrium
on the nightside. In principle, some hot exoplanets may be
preferentially unstable near their limb, which will be an interesting
issue to explore with improved instability models.

\begin{figure*}[l]
\centering \includegraphics[scale=0.8]{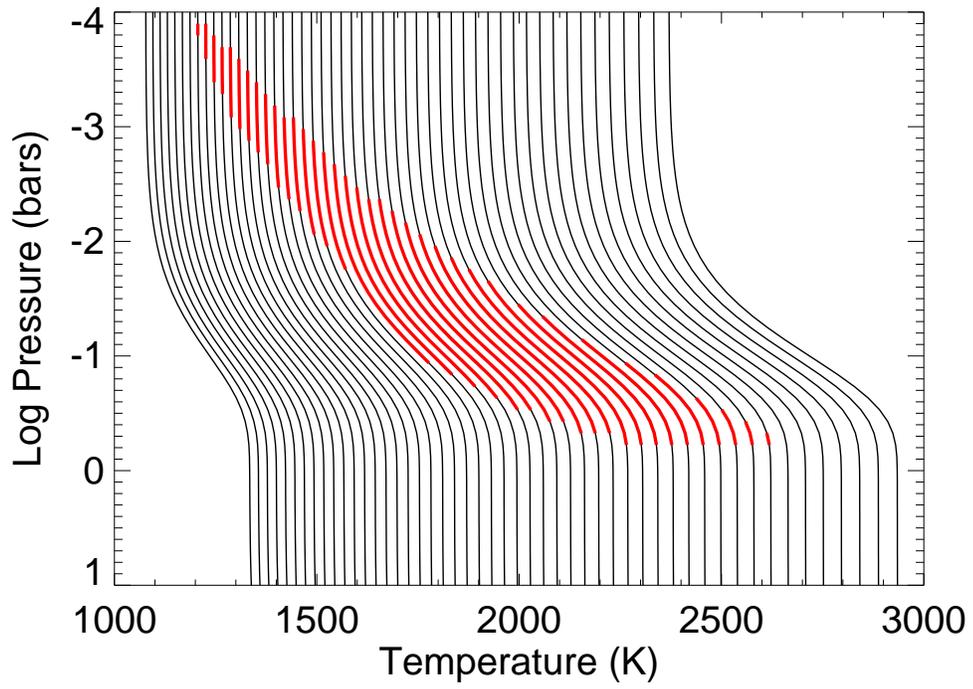}
\caption{Same as Figure~1 with a weaker magnetic drag law ($n_s=0.3$,
  $n_e=1$ in Eq.~[\ref{eq:drag}]), $V_\phi=7$~km~s$^{-1}$ and $B=10$~G. This
  drag law delays the transition into the strong drag regime to higher
  temperatures, which extends the domain of instability (compare to
  the green domain in Figure~1).}
\label{fig:two}
\end{figure*}

\begin{figure*}[l]
\centering \includegraphics[scale=0.8]{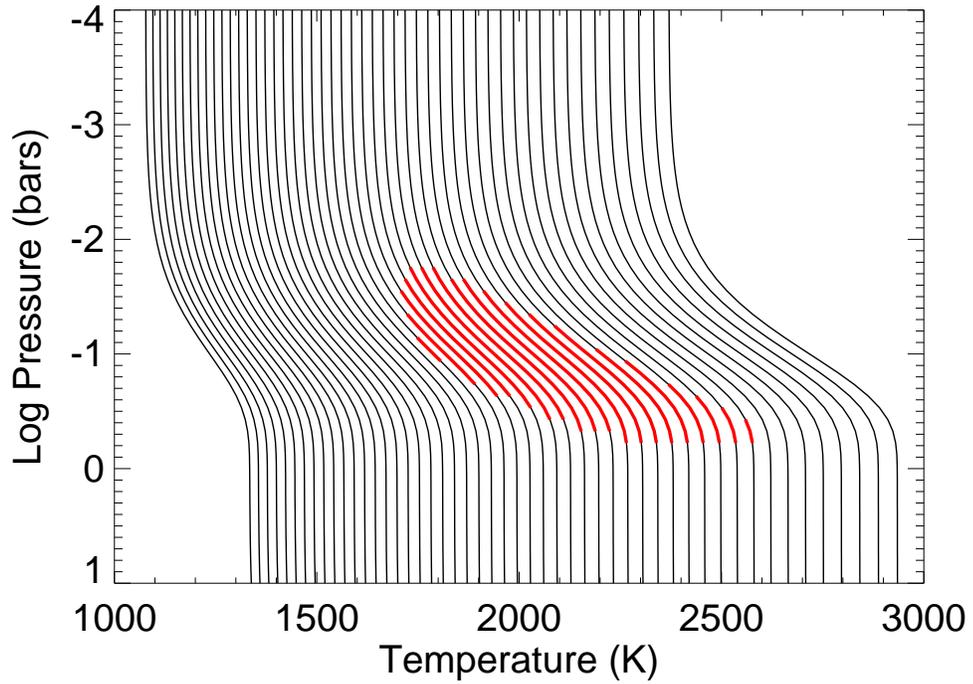}
\caption{Same as Figure~2 with a constant ionization fraction
  $x_e^+=10^{-8}$ added throughout the atmosphere, to mimick the
  effects of photoionization. The additional, constant ionization
  contribution stabilizes the upper atmospheric regions by inducing a
  stronger drag regime. The entire atmospheric domain can be
  stabilized with the addition of a constant ionization fraction
  $x_e^+ \sim 10^{-7}$.}
\label{fig:three}
\end{figure*}

The outcome of the instability is difficult to anticipate beyond
qualitative expectations. Unstable regions should reach temperatures
high enough for saturation of the instability in the strong drag
regime, which can amount to temperature excesses of several hundreds
Kelvin according to the width of domains shown in
Figs.~\ref{fig:one}--\ref{fig:three}. Perhaps more importantly, the
radiative response of the vertically-coupled atmospheric layers,
together with the horizontal coupling caused by locally modified
thermal and drag conditions, will result in a very non-linear response
that is best studied with global models. While short instability
growth times may prove numerically challenging, global circulation
models with adequate MHD treatments offer a promising avenue for
progress. They would help clarify the global energetics of the
instability, which is ultimately powered by the same differential
thermal forcing as the global circulation itself.

\acknowledgments 

The author thanks E. Rauscher for comments on the manuscript.  This
work was supported in part by NASA grant PATM NNX11AD65G.

\newpage

\end{document}